\documentclass[%
superscriptaddress,
 amsmath,amssymb,
 aps,
]{revtex4-1}

\usepackage{graphicx}% Include figure files
\usepackage{dcolumn}% Align table columns on decimal point
\usepackage{bm}% bold math
\usepackage{xcolor}% Allows for color highlighting of comments

\begin{document}

\title{Real time $g^{(2)}$ monitoring at 100\,kHz}

\author{Carolin L\"uders}
\author{Johannes Thewes}
\author{Marc A{\ss}mann}

\affiliation{Experimentelle Physik 2, Technische Universit\"at Dortmund, 44227 Dortmund, Germany}


\date{\today}

\begin{abstract}
We introduce a technique to determine photon correlations of optical light fields in real time. The method is based on ultrafast phase-randomized homodyne detection and allows us to follow the temporal evolution of the second-order correlation function $g^{(2)}(0)$ of a light field. We demonstrate the capabilities of our approach by applying it to a laser diode operated in the threshold region. In particular, we are able to monitor the emission dynamics of the diode switching back and forth between lasing and spontaneous emission with a $g^{(2)}(0)$-sampling rate of 100\,kHz.  
\end{abstract}

%\ocis{(270.5290) Photon Statistics; (120.2920) Homodyning.} % REPLACE WITH CORRECT OCIS CODES FOR YOUR ARTICLE, MINIMUM OF TWO; Avoid using the OCIS codes for “General” or “General science” whenever possible.
%For a complete list of OCIS codes, visit: https://www.osapublishing.org/oe/submit/ocis/
\maketitle
%%%%%%%%%%%%%%%%%%%%%%% References %%%%%%%%%%%%%%%%%%%%%%%%%

%%%%%%%%%%%%%%%%%%%%%%%%%%  body  %%%%%%%%%%%%%%%%%%%%%%%%%%
\section{Introduction}
Temporal coherence properties and photon statistics of optical light fields are widely studied in order to gain insights about the characteristics of light sources and the dynamics of the light field. They have been utilized in different fields such as fluorescence correlation spectroscopy in chemistry \cite{Magde1972}, identifying single photon emission via antibunching \cite{Kimble1977} or evaluating the position of the laser threshold in high-$\beta$ lasers \cite{Rice1994}. An adequate measurement of the coherence properties of light requires an experimental setup with a temporal resolution at least comparable to the coherence time of the light field under investigation.\\
Since the first demonstrations of photon bunching using two photomultipliers \cite{HBT1956}, detector technology has advanced significantly in terms of sensitivity and quantum efficiency\cite{Stevens2010,Zhou2013,Moody2018}. Also, several experimental techniques have been developed in order to allow for studies of photon correlations at short timescales. These include using processes that are directly sensitive to the presence of several photons such as two-photon absorption \cite{Boitier2009} or upconversion \cite{Hayat2010} or different detector architectures, such as streak cameras \cite{Assmann2009,Assmann2010} or time multiplexed detection \cite{Avenhaus2010}. These approaches focused on increasing the temporal resolution in order to adequately study the second-order correlation function $g^{(2)}(\tau)$ for light fields with short coherence times. $g^{(2)}(\tau)$ can be interpreted as the conditional probability to detect a second photon at a delay $\tau$ with respect to the detection of a first photon. Considering the equal-time correlation function $g^{(2)}(\tau=0)$, one may distinguish between different kinds of light fields. For example, coherent light fields show uncorrelated photon emission resulting in $g^{(2)}(0)=1$, while thermal light fields show photon bunching, which corresponds to $g^{(2)}(0)=2$ \cite{Glauber1963}.\\
For single-mode emitters, knowledge about $g^{(2)}(\tau)$ is fully sufficient for a thorough characterization of the emitted mode of interest. For multimode emitters, however, the underlying physics may become much more complicated as a complex interplay between different modes of the light field and the emitter may arise. As a consequence, non-trivial mode dynamics may occur and $g^{(2)}(\tau)$ will not necessarily be stationary, but may instead acquire an explicit time-dependence, which is known from effects such as mode-hopping in lasers \cite{Gray1991}. Identifying such effects is a challenging task because time averaged measurements of $g^{(2)}(0)$ cannot distinguish whether the measured value of $g^{(2)}(0)$ is constant in time or a weighted average of a time-dependent $g^{(2)}(0)$. In order to differentiate between these two cases, one needs an experimental technique that simultaneously features a sufficient temporal resolution to measure $g^{(2)}(0)$ and a fast sampling rate that is able to capture its dynamics.\\
In this work, we establish such an experimental technique tailored towards multimode light fields, which may show a non-stationary value of $g^{2}(0)$ that varies with time. We demonstrate the capabilities of our setup by monitoring the coherence properties and bistable emission of an external-cavity diode laser across the threshold region. Our approach allows us to continuously monitor the coherence properties of the emission with a temporal resolution of 120\,fs, while simultaneously achieving a $g^{(2)}$-sampling rate of 100\,kHz corresponding to a sampling time of 10\,$\mu$ s per data point, which opens up the possibility to perform real-time monitoring of $g^{(2)}(0)$. 

\section{Time-resolved photon correlations}
The central quantity of interest in characterizations of non-stationary light fields is the second-order correlation function, which is defined as:
\begin{equation}
g^{2}(\tau,t)=\frac{\langle \hat{a}^\dagger(t) \hat{a}^\dagger(t+\tau) \hat{a}(t+\tau) \hat{a}(t)\rangle}{\langle \hat{a}^\dagger(t) \hat{a}(t)\rangle \langle \hat{a}^\dagger(t+\tau) \hat{a}(t+\tau)\rangle },
\label{eq:g2}
\end{equation}
where the operators denote bosonic creation and annihilation operators for the mode of interest of the light field.\\
It is worthwhile to emphasize that $g^{2}(\tau,t)$ features two relevant timescales. The first one refers to the delay $\tau$. For any stationary light field, $g^{(2)}(\tau)$ will return to the value of unity for delays long compared to the coherence time of the light field. Since the temporal resolution $\Delta \tau$ of the experimental setup determines the meaningful resolution in $\tau$, it should at least match the coherence time $\tau_c$ of the light field. Otherwise, if $\tau \gg \tau_c$, the experimental results will reflect a convolution of the real $g^{2}(\tau, t)$ with the instrument response function of the detector \cite{Ulrich2007}.\\
The second important timescale corresponds to $t$ and describes the dynamics of the second-order correlation function. The meaning of this timescale depends on the evaluation of the expectation values in equation (\ref{eq:g2}), which need to be interpreted differently for pulsed and continuous wave light fields. For pulsed fields, they may be considered as an ensemble average over a large number of identically prepared pulses and one may investigate the progression of $g^{(2)}(0,t)$ during the course of the emission pulse \cite{Adiyatullin2015,Assmann2010a}. Usually, $\Delta t \approx \Delta \tau$ in such cases.\\
For continuous wave emission one may interpret these expectation values as time averages and $\Delta t$ corresponds to the sampling time per $g^{(2)}(0)$ value. For example in our experiment one photon number measurement with a temporal resolution of $\Delta \tau = 120\,$fs is performed every 13.3\,ns. Evaluating the expectation values of 754 consecutive measurements results in a sampling time of 10\,$\mu$ s per data point, which is sufficient to reduce statistical errors in $g^{(2)}(0)$ to an acceptable amount. In that case, dynamics in the coherence properties of the light field can still be resolved on a timescale of $\Delta t = 10$\,$\mu$ s while maintaining the intrinsically better temporal resolution of $\Delta \tau =$ 120\,fs for the photon number measurement. As a non-constant value of $g^{(2)}(0,t)$ indicates a multimode state of the light field \cite{Avenhaus2010}, its dynamics are relevant for pulsed light fields as well as inherently multimode continuous wave emitters such as bimodal lasers with mode competition. Experimental techniques based on time-resolved single photon counting typically suffer from pile-up and detector dead times\cite{Arlt2013} which result in integration times on the order of seconds. Accordingly, although $g^{(2)}(0,t)$ in principle contains information about mode competition and other intrinsically multimode effects, only dynamics on very slow timescales are captured by these experiments.\\
Instead of direct photon counting approaches as outlined above, also continuous variable strategies for measuring the photon correlation functions of light fields have been demonstrated using balanced homodyne detection \cite{Roumpos2013a}. Balanced homodyne detection is a powerful tool in continuous variable quantum optics as it provides direct experimental access to the field quadratures 
\begin{align}
	\hat{q}&=\frac{1}{\sqrt{2}} (e^{i \phi} \hat{a}^\dagger + e^{-i \phi} \hat{a}^{})
\end{align}
of a mode of the light field at phase $\phi$. Its applications range from quantum state tomography of light fields using phase-resolved balanced homodyne detection \cite{Lvovsky2009} up to phase-averaged balanced homodyne detection for ultrafast measurements of photon statistics limited in temporal resolution only by the duration of the reference local oscillator pulse \cite{Munroe1995,McAlister1997}. The fast timescales accessible using homodyne detection opened up the path towards investigations of pulsed laser dynamics on the sub-ps scale \cite{Blansett2001} performed on an ensemble of laser pulses prepared identically at kHz repetition rates\cite{Roumpos2013}.\\
In a homodyne detection experiment, the field quadratures of a weak signal field are obtained in three steps \cite{Roumpos2013,Lvovsky2009}. First, the signal light field (S) interferes on a beamsplitter with the local oscillator (LO), a strong coherent light field, which yields the sum and difference of field operators in the two exit ports of the beamsplitter, respectively. Second, the two output beams are detected separately by two photodiodes. The detection operation on the sum ($+$) and difference ($-$) photodiode can be expressed by a combined operator $\hat{a}_{\pm}$ constructed by bosonic annihilators $\hat{a}_\text{LO}$ for the local oscillator and $\hat{a}_{S}$ for the signal:
\begin{align}
	\hat{a}_{\pm} = \frac{1}{\sqrt{2}} \left( e^{i \phi} \hat{a}_{\text{LO}} \pm \hat{a}_{\text{S}} \right)
\end{align}
Here, $\phi$ denotes the relative phase between LO and S. In the third step, the difference photocurrent is recorded, which is given by $\hat Q = \hat{a}^\dagger_+ \hat{a}^{}_+ - \hat{a}^\dagger_- \hat{a}^{}_-$. Assuming a strong and coherent LO, we can separate the contribution of bosonic annihilation and creation operators and arrive at
\begin{align}
	\hat Q = \langle \hat{a}_{\text{LO}} \rangle \left( e^{i \phi} \hat{a}^\dagger_{\text{S}} + e^{- i \phi} \hat{a}^{}_{\text{S}} \right).
\end{align}
This term is directly proportional to the field quadrature $\hat{q}$ of the signal light field. For pulsed local oscillator light fields, one measurement of the operator $\hat{Q}$ is performed for each LO pulse and we obtain a series $\hat{Q}(t)$ of measurements discretized in time.\\
We can also calculate the equal-time second-order correlation function
\begin{align} \label{eqn:equalTimeg2}
	g^{(2)}(\tau = 0, t) = \frac{\langle \hat{a}^\dagger_{\text{S}} \hat{a}^\dagger_{\text{S}} \hat{a}^{}_{\text{S}} \hat{a}^{}_{\text{S}} \rangle}{{\langle \hat{a}^\dagger_{\text{S}} \hat{a}^{}_{\text{S}} \rangle}^2}
\end{align}
from the moments $\langle \hat{Q}^2 \rangle$ and $\langle \hat{Q}^4 \rangle$ of $\hat{Q}$\cite{Roumpos2013}. Note that we omit the time dependencies of $\hat Q$, $\hat{a}^\dagger_\text{S}$ and $\hat{a}^{}_\text{S}$ for readability. The calculation of $g^{(2)}(0)$ is possible because $\langle \hat{Q}^2 \rangle$ includes the denominator of equation (\ref{eqn:equalTimeg2}) and $\langle \hat{Q}^4 \rangle$ its numerator:
\begin{align}
	\langle \hat{Q}^2 \rangle &= \langle \hat{a}^{}_{\text{LO}} \rangle^2 \left( 2 \langle \hat{a}^\dagger_{\text{S}} \hat{a}^{}_{\text{S}} \rangle + 1\right) \label{eqn:Q2}\\
	\langle \hat{Q}^4 \rangle &= \langle \hat{a}^{}_{\text{LO}} \rangle^4 \left( 6\langle \hat{a}^\dagger_{\text{S}} \hat{a}^\dagger_{\text{S}} \hat{a}^{}_{\text{S}} \hat{a}^{}_{\text{S}} \rangle + 12\langle \hat{a}^\dagger_{\text{S}} \hat{a}^{}_{\text{S}} \rangle + 3 \right) \label{eqn:Q4}
\end{align}
To compute the moments in equations (\ref{eqn:Q2}) and (\ref{eqn:Q4}), it is essential to assume phase averaged measurements. Then, the expectation value of each term containing a phase dependence $e^{i n \phi}$ vanishes and we can use $\langle \hat Q \rangle = 0$. In the end, we can obtain the equal-time second-order correlation function $g^{(2)}(0,t)$ of the signal light field from a series of measurements of $\hat{Q}(t)$ by evaluating the above expectation values for a certain number of local oscillator pulses.\\
For homodyne detection, the data acquisition rate is given by the repetition rate of the laser used as the local oscillator. Real-time monitoring lays high demands on the experimental setup. As outlined above, uniform sampling of all relative phases $\phi$ between the signal field and the local oscillator is a fundamental requirement for computing $g^{(2)}(0)$. Accordingly, the repetition rate of the local oscillator and the bandwidth of the detector are of prime importance to achieve real-time $g^{(2)}(0)$ sampling of the light field. Recently, wideband balanced detector architectures suitable for operation at standard Ti:Sapphire laser repetition rates up to 100 MHz have been introduced \cite{Kumar2012}. In the next section, we will describe an experimental setup that utilizes this detector architecture and opens up the possibility for measuring $g^{(2)}(0,t)$ at sampling rates of 100\,kHz.

\section{Experimental Setup}
\begin{figure}[ht!]
\centering\includegraphics[width=0.7\textwidth]{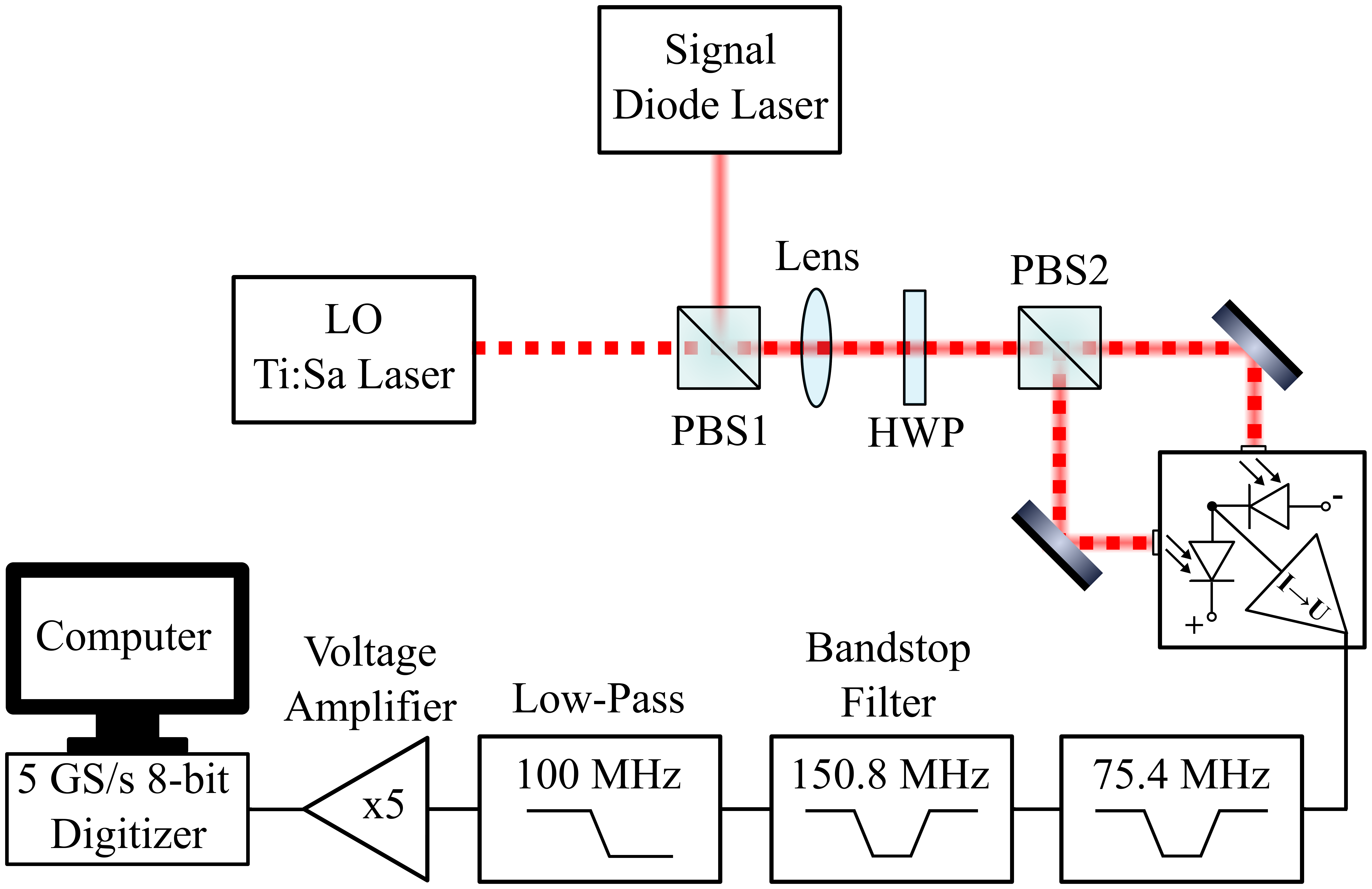}
\caption{Schematic of the homodyne detection setup. LO: local oscillator; PBS: polarizing beam splitter; HWP: half-wave plate.}
\label{fig:setup}
\end{figure}

Our homodyne detection setup is shown schematically in Figure \ref{fig:setup}. 
The local oscillator (LO), which is a strong coherent light field, and the signal are combined on a polarizing beam splitter (PBS1). After this a half-wave plate (HWP) can be adjusted in order to divide the combined beam on the second polarizing beam splitter (PBS2) such that the two photodiodes deliver an equal average photocurrent. Then the difference photocurrent is amplified and digitized. Whereas its average is zero, its fluctuations provide a measurement of the field quadrature $\hat{q}$ of the signal amplified by the amplitude $\langle \hat{a}_{LO} \rangle$ of the LO. Hereby only those field modes that overlap spectrally, spatially and temporally with the LO are amplified. From the recorded field quadratures, the second-order correlation function $g^{(2)}(0)$ may then be calculated by evaluating the statistical moments of a sufficient number of samples according to equations (\ref{eqn:equalTimeg2}-\ref{eqn:Q4}).\\
The time resolution of this experiment is determined by three numbers: First, the temporal resolution $\Delta \tau$ is given by the duration of one LO pulse, as during this time the signal photons are amplified. Our local oscillator pulses are provided by a Ti:sapphire laser and have a duration of about 120\,fs. Second, the sampling rate with which the field quadratures are recorded corresponds to the LO repetition rate of 75.4\,MHz. Third, the sampling time per $g^{(2)}(0)$ value depends on how many quadrature samples are used for accumulating statistics. By using 754 samples, we end up with a $g^{(2)}(0)$ sampling rate of 100\,kHz corresponding to a sampling time of $\Delta t = 10$\,$\mu$ s per value of $g^{(2)}(0)$.\\
These are the fundamental time scales given by the LO itself. We ensure that also the components of the detection and data acquisition setup provide sufficient bandwidth. For the detection we use a commercial balanced homodyne detector with a bandwidth of 100\,MHz provided by FEMTO (Model HCA-S). It contains two Si-PIN photodiodes and a transimpedance amplifier with a gain of 5\,kV/A. We verified shot noise limited operation for the LO power of 5\,mW we employ. \\
The LO repetition rate and its higher harmonics are removed from the signal by notch filters at 75.4\,MHz and 150.8\,MHz from Rittmann-HF-Technik and a 100\,MHz low-pass filter from Crystek. The filtered signal is then amplified by a Stanford Research Systems SR445 preamplifier with a bandwidth of 300\,MHz. For data acquisition a M4i.2234-x8 digitizer from Spectrum is used providing sampling rates of 5\,GS/s at 8\,bit resolution. The acquired voltages are then numerically integrated for each LO pulse in order to determine the number of photoelectrons and from this the quadrature value.\\
The signal we investigate comes from a Toptica DL pro Littrow-type external-cavity diode laser emitting continuously at a central wavelength of 835\,nm. In order to increase the lasing threshold and to introduce non-trivial dynamics of the laser mode coherence, we deliberately misaligned the feedback slightly, so that the laser does not show ideal single mode operation. For a single drive current, a whole run of the experiment consists of 16 million consecutive quadrature measurements, which are divided into 21455 batches of 754 measurements each, which corresponds to a sampling time of $\Delta t=$10\,$\mu$ s. Single values of the mean photon number and $g^{(2)}(0)$ are calculated for each batch.\\

\section{Results}
\begin{figure}[ht!]
\centering\includegraphics[width=0.8\textwidth]{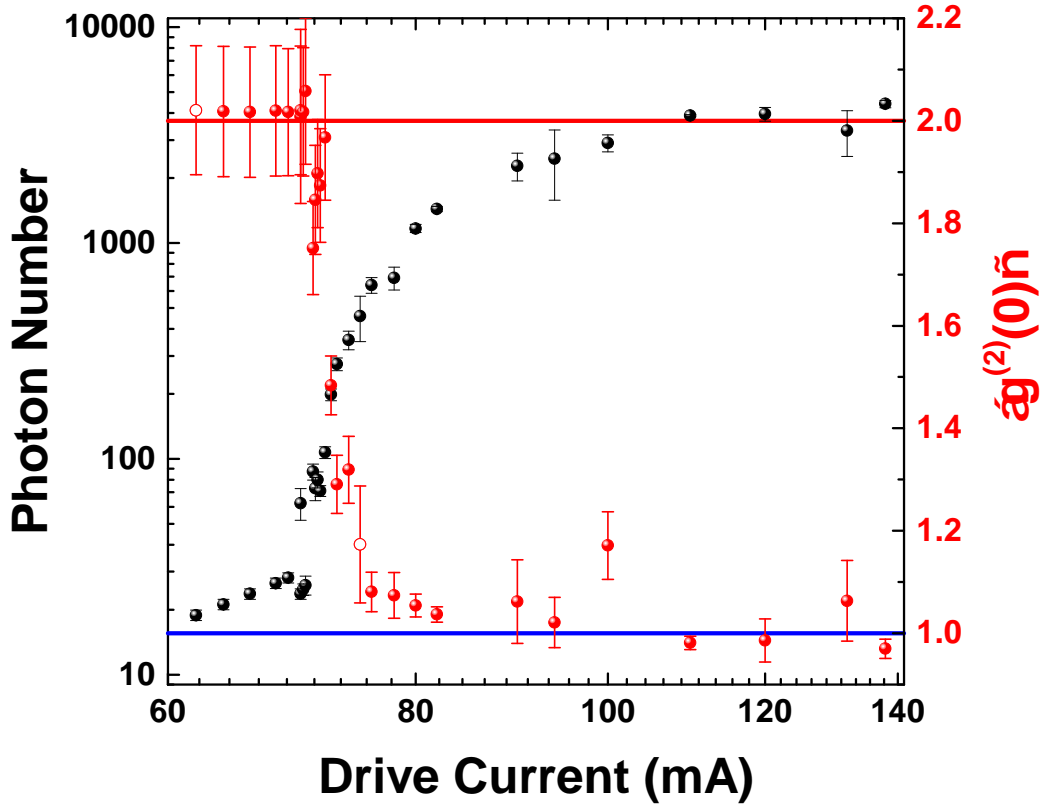}
\caption{Time-averaged emitted photon number (black dots) and second-order correlation function $\langle g^{(2)}(0) \rangle$ (red dots) plotted against the drive current across the lasing threshold region. Red and blue solid lines denote the thermal and coherent limit of $\langle g^{(2)}(0) \rangle$, respectively. The open red dots mark the drive currents investigated in more detail in figure \ref{fig:g2vsn}.}
\label{fig:IO}
\end{figure}

\begin{figure}[ht!]
\centering\includegraphics[width=0.8\textwidth]{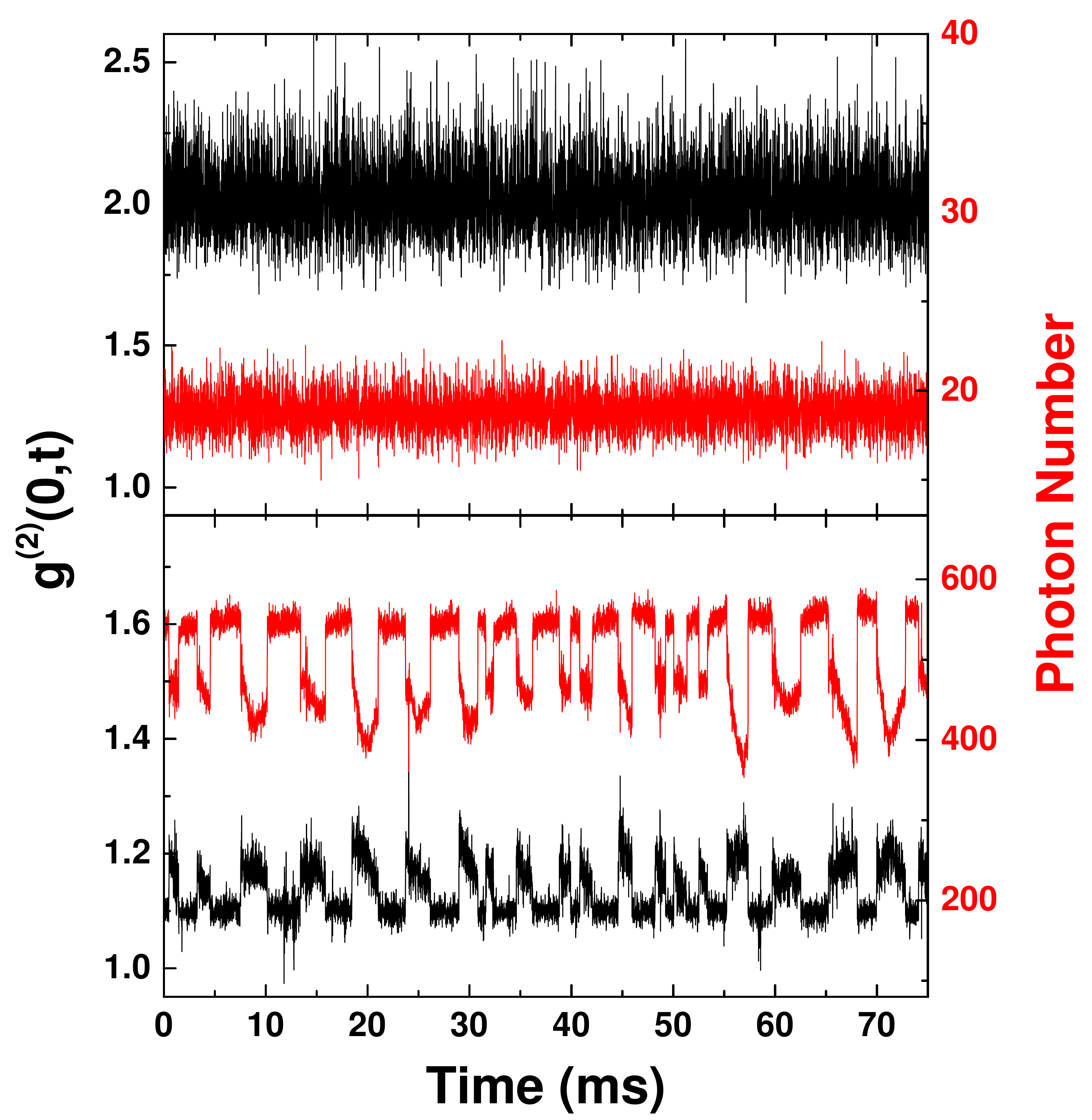}
\caption{Simultaneously recorded time trace of the emitted photon number and $g^{(2)}(0)$ for different drive currents ($I_t$=70.5\,mA). Their dynamics clearly show correlated fluctuations. From top to bottom, the currents used are 62\,mA and 75\,mA, respectively. }
\label{fig:g2vsn}
\end{figure}

\begin{figure}[ht!]
\centering\includegraphics[width=\textwidth]{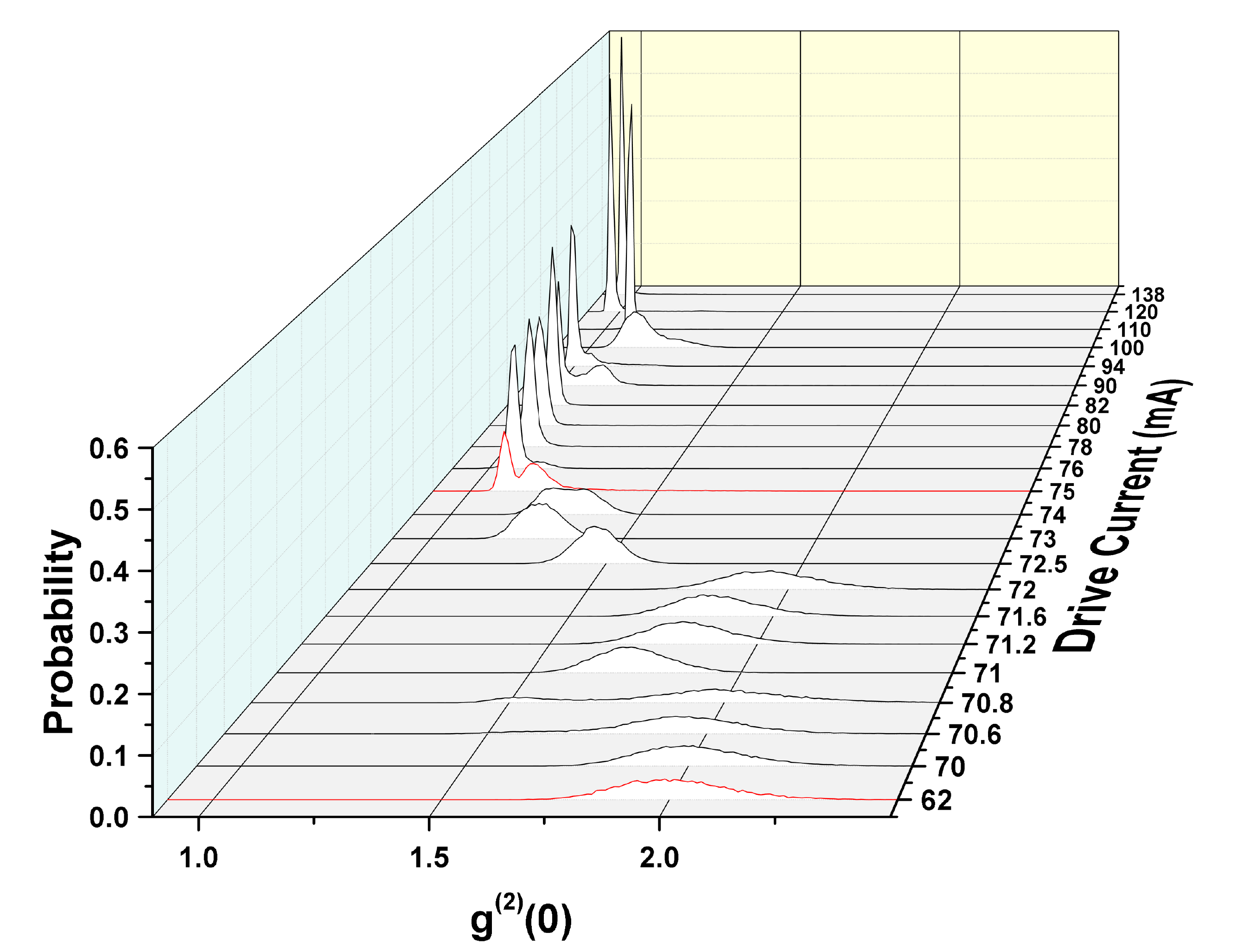}
\caption{Histograms of the distribution of the values of the second-order correlation function for different drive currents. Note that the spacings of drive currents along the z-axis are not equidistant. The datasets used for figure \ref{fig:g2vsn} are marked in red.}
\label{fig:g2hist}
\end{figure}

As an example of the capabilities of our ultrafast $g^{(2)}$-monitor, we have investigated the dynamics of the external cavity diode laser introduced in the last section. At the lasing threshold we expect to see two different effects. First, a steep rise in the emitted number of photons characterizes the lasing threshold. Second, we should see thermal light with almost exactly $g^{(2)}(0)=2$ below this lasing threshold and coherent light with almost exactly $g^{(2)}(0)=1$ above it. Figure \ref{fig:IO} shows the input-output curve in black dots and the mean value of the second-order correlation function $\langle g^{(2)}(0)\rangle$ in red ones. Here, the results represent averages over the values for all batches and the error bars correspond to their standard deviation. As expected, the results show the typical transition from the emission of thermal light below threshold towards lasing emission far above the threshold. The transition takes place at a threshold current of about $I_t$=70.5\,mA and is associated with a steep rise of the emitted photon number.\\
It is striking that several data points show comparably large variations in $g^{(2)}(0)$ as denoted by the error bars in figure \ref{fig:IO}. For example at a drive current of 75\,mA, the spread of $g^{(2)}(0)$ is almost as large as below threshold, at 62\,mA, although the mean photon number differs already by more than one order of magnitude for these drive currents. Unfortunately, we are not able to identify the origin behind this and other enhanced error bars by looking at the averaged values $\langle g^{(2)}(0)\rangle$ alone. They could, for example, arise due to imprecision in the measurements or due to some physical origin such as multimode behavior. Therefore, we investigate the lasing threshold more thoroughly by studying not only the averaged, but also the time-resolved correlation function $g^{(2)}(0,t)$ with a sampling time of $\Delta t =$ 10\,$\mu$ s.\\
In order to learn more about the origin of unexpectedly large error bars in figure \ref{fig:IO}, we investigate the results for the drive currents of 75\,mA and 62\,mA in more detail. Figure \ref{fig:g2vsn} shows time traces of both the emitted photon number and $g^{(2)}(0)$ of the emission for these drive currents.
At 62\,mA $g^{(2)}(0)$ shows large scale fluctuations around the value of 2 expected for the thermal regime. For any drive current, we carefully ensured that the sampling time $\Delta t$ per $g^{(2)}(0)$ value is several orders of magnitude larger than the coherence time of the emitted light and checked that the fluctuations become smaller as larger values for $\Delta t$ are chosen (data not shown). Therefore, the deviations from the expected value of 2 below threshold can be fully explained by statistical noise.\\
On the contrary, the time trace for the drive current of 75\,mA immediately makes it clear that the spread of the values of $g^{(2)}(0)$ in this case is not only due to statistical noise, but has a physical origin. Both, the photon number and $g^{(2)}(0)$ move back and forth between two rather well defined states: The photon number is mostly either close to 450 or to 550 photons and $g^{(2)}(0)$ is either close to 1.1 or to 1.2. Other values are taken on only rarely. These fluctuations are also strongly correlated. The state with the higher photon number correlates with the lower $g^{(2)}(0)$ and vice versa. This bistable profile of the emitted intensity is well known for semiconductor lasers subject to mode hopping \cite{Gray1991} and it is interesting to note that the second-order coherence of the emission shares this bistable behaviour. Still, especially close to switching points the correlations between the photon number and the coherence properties of the emission do not seem to be exact. Accordingly, a detailed study of both quantities may provide deeper insights into the mode hopping dynamics, which is, however, out of the scope of the present work.\\
While the time traces presented above provide an intuitive understanding of the effects taking place during the lasing transition, it is desirable to achieve a more quantitative description of the additional information we gain about the diode laser coherence properties across the threshold. We follow a statistical approach and calculate histograms of the relative frequencies of the individual values of $g^{(2)}(0,t)$. Figure \ref{fig:g2hist} shows a set of these histograms for different drive currents. While time-integrated experiments yield only the weighted average value, histograms provide full access to the probability distribution of $g^{(2)}(0,t)$ values.
\\
We start our analysis with the two drive currents of 62\,mA and 75\,mA which we already investigated in the time domain. Below threshold, for 62\,mA, the distribution has a broad Gaussian shape and is centered on $g^{(2)}(0)$ = 2 as expected for thermal light. When looking at the probability distribution for 75\,mA, the picture changes drastically. Instead of a single Gaussian peak, we find two with the first one being narrower than the second. This feature corresponds to the bistability already found in the time traces. With the histogram, however, we gain access to the $g^{(2)}(0)$ of the individual modes and their relative strength. Using this tool, we are now able to analyze the $g^{(2)}(0)$ values from figure \ref{fig:IO} and their error bars.\\
When we have a closer look at the drive current of 70.8\,mA, we see that the emission becomes bistable for the first time as evidenced by a weak contribution showing $g^{(2)}(0)$ = 1.5 and a much stronger contribution that appears to be fully thermal. Increasing the drive current from there within the narrow window of values between 71\,mA and 72\,mA the bistability vanishes, but surprisingly $g^{(2)}(0)$ increases from 1.75 up to 1.9 again. One possible explanation might be that the bistability seen beforehand can be attributed to two coexisting modes which are mostly independent of each other. In that case, photons from the two modes are uncorrelated. Thus, the value of $g^{(2)}(0)$ of the total emission will be reduced significantly if the two modes show comparable emission intensity. Accordingly, the increase in $g^{(2)}(0)$ may be interpreted as the result of the emission of two modes. While they are of roughly similar intensity for the lower drive currents around 71\,mA, one of the modes becomes dominant at higher drive currents up to 72\,mA. Within this region $g^{(2)}(0)$ of the total emission increases towards the value of $g^{(2)}(0)$ for the stronger mode as the uncorrelated emission from other modes decreases. Within the transition region stable operation and bistability alternate, as can be seen for example for a drive current of 75\,mA, where the bistability emerges clearly. Increasing the drive current even further, the emission quickly reaches the standard lasing regime at a drive current of about 80\,mA. Moving to even higher drive currents, the diode laser mostly shows stable lasing with few regions of instability appearing for selected values of drive currents. Still, at high drive currents the lasing process is dominantly stable as can be deduced from the small width of the distribution of values of $g^{(2)}(0)$ at a drive current of 138\,mA, which shows that analysis of the full histograms of the distribution of $g^{(2)}(0)$ values is an adequate tool to characterize the dynamics of a laser mode.

\section{Conclusion}
In summary, we have presented an experimental technique that opens up the possibility to perform real time measurements of photon correlations. The intrinsic temporal resolution of the quadrature measurement is given by the duration of the local oscillator which can be on the order of 100\,fs or even shorter and the sampling time required for a single value of $g^{2}(0)$ amounts to about 10\,$\mu$s. We have demonstrated the capabilities of our setup by investigating an external cavity diode laser in the threshold region and have been able to show that the emitted photon number and degree of second-order correlation show fluctuations that are synchronized with each other. We envision that our technique will be valuable in advancing the understanding of nanolasers \cite{Wang2015} and time-resolved photon correlations \cite{Valle2012} especially with respect to feedback \cite{Albert2011} and to investigate phenomena that occur spontaneously and do not yield a signal to trigger on, such as some kinds of superradiance \cite{Jahnke2016}. Further, our setup can be extended to more than one homodyne detection channel, which will allow us to study several modes simultaneously. This opens up the path towards real time studies of mode competition in bimodal lasers \cite{Marconi2018,Leymann2013, Leymann2017} and the mode switching behavior associated with it \cite{Redlich2016}. Instead, one could also use several homodyne detection channels in order to study the same mode. By placing the local oscillators at different relative delays, the effective sampling time per $g^{(2)}(0)$ value can be reduced n-fold by using n detection channels. Also, it should be noted that the sampling time of 10\,$\mu$ s that we use is not fixed. It is always possible to reduce the sampling time if broader histograms of $g^{(2)}(0,t)$ are acceptable and vice versa. 

\section*{Funding}
This work was supported by the German Science Foundation (DFG) in the framework of grant AS 459/1-2.

\section*{Acknowledgments}
We thank Alexander Lvovsky for helpful discussions about high bandwidth homodyne detection.

%\bibliographystyle{osajnl}
%\bibliography{RealTimeg2}
\bibliography{RealTimeg2}

\end{document}